\documentstyle[preprint,aps]{revtex}
\begin{document}
\title {\bf Reconstruction and thermal stability
            of the cubic SiC (001) surfaces}
\author{Alessandra Catellani$^{(a)}$, Giulia Galli$^{(b)}$ and 
Fran\c{c}ois Gygi$^{(b)}$}
\address{$(a)$~CNR-MASPEC, Via Chiavari, 18/A, I-43001 Parma, Italy} 
\address{
$(b)$~
 Institut Romand de Recherche Num\'erique
 en Physique des Mat\'eriaux (IRRMA)
 CH-1015 Lausanne, Switzerland
}
\maketitle
\begin{abstract}
The (001) surfaces of cubic SiC were investigated with ab-initio
molecular dynamics simulations.
We show that C-terminated surfaces can
have different c(2$\times$2) and p(2$\times$1) reconstructions,
depending on preparation conditions and
thermal treatment, and we suggest experimental probes 
to identify the various reconstructed geometries.
Furthermore we show 
that Si-terminated surfaces exhibit a p(2$\times$1) reconstruction
at $T=0$, whereas above room temperature they oscillate 
between a dimer row and an ideal geometry below 500 K, and sample
several patterns including a c(4$\times$2)  above 500 K.
\end{abstract}

 Silicon Carbide is an attractive material for high temperature micro-
and optoelectronic devices\cite{hmss94} because of 
its wide band gap, high thermal
conductivity, high hardness and chemical inertness. 
Having a small lattice mismatch with GaN,
SiC has also emerged as a promising substrate for the growth of 
nitride-based devices\cite{hmss94,depn96}.

 In the last decade, a notable effort has been devoted to the characterization 
of SiC surfaces, since SiC films for use in technological applications are 
prepared by epitaxial growth.
In particular, the (001) surfaces of the cubic polytype $\beta$-SiC
have been studied with a variety 
of experimental [3-11]
and theoretical [12-15] techniques.
At the end of the eighties, it was 
established that these surfaces are terminated by only one
species and a clear assignment of
different LEED patterns to either C- or Si-terminated
surfaces was given\cite{k89}. 
Nevertheless the reconstruction of
the Si-terminated surface is still the subject of
debate, in view of the partial disagreement
between recent\cite{s96} and older\cite{k89,pmhs92} experiments.
Furthermore, 
a long-standing controversy\cite{bk91,ppmhs91,cs911,b91,ysj95} about the
structure of C-terminated surfaces was solved only recently\cite{lbr96}.
At present, important problems such 
as the effect of preparation conditions on 
the surface structure and the thermal stability of the (001) surfaces 
are yet unsolved. These are key issues for the understanding of
any growth process on SiC substrates, 
and represent as well fundamental problems
in the physics of compound semiconductor surfaces. 

 In order to address these issues, 
we performed a series of ab-initio
molecular dynamics (MD) simulations\cite{CP85} of several
$\beta$-SiC (001) surfaces at finite temperature. We studied
various reconstruction paths of the C- and Si-terminated surfaces, 
modeling different preparation conditions, and we
identified experimental probes which could discriminate between the
various reconstruction patterns.
Our computations constitute the first ab-initio analysis of
the effect of preparation conditions and thermal treatmtent on 
the structure of (001) SiC surfaces.

 Our calculations were carried out within the local density functional
approximation (LDA).  We used slabs periodically
repeated in the (x,y) plane (see Fig.~1), 
terminated on each side by the same atomic species and 
followed by a vacuum region of $\simeq$ 8.5 \AA.
In most MD simulations we used 11 layers, each containing 8 atoms.
Structural optimizations at $T=0$ were carried out 
both with 11 and 19 layers (152 atoms)\cite{convergence}. 
The lateral dimensions of the supercell were 8.60 $\times$ 8.60 \AA, 
corresponding to the theoretical equilibrium lattice 
constant of bulk SiC (4.30 \AA). The experimental value is 4.36 \AA.
We note that the use of a symmetric slab is essential to avoid 
spurious charge transfers
and electric fields, which could be present in H-terminated 
non symmetric slabs\cite{skmrp96}, 
or in slabs terminated by different atomic species on 
opposite sides\cite{ysj95}.
The interaction between ionic cores and valence
electrons was described by fully non-local pseudopotentials\cite{drh89}
 with s and p non locality for Si and s-only non locality for C.
 Single particle orbitals and charge densities were expanded
in plane waves  with kinetic energy cutoffs ($E_{\rm{cut}}$) of 40 and 160 Ry, 
respectively.
We considered Bloch functions at the $\overline{\Gamma}$ point
of the supercell surface Brillouin zone (SBZ); this corresponds to
including the points
$\overline{\Gamma}$, $\overline{M}$, $\overline{S}$,
and $\overline{\Gamma}$, $\overline{J'}$ and 
($\overline{J}$+$\overline{J'}$/2) of the c(2$\times$2) and p(2$\times$1)
primitive cells (001) SBZ, respectively.

{\it C-terminated surfaces} --
 Experimentally, C-terminated surfaces are obtained either
by ethylene decomposition on Si-terminated substrates\cite{bk91,ppmhs91},
or by Si sublimation\cite{k89,pc91,ssmc95} between 1200 and 1500 K.
The two techniques lead to samples characterized by 
different diffraction data\cite{ppmhs91} and 
photoemission spectra\cite{pc91,bl95}.
We devised two series of computations, to investigate the two
preparation methods.

 The first series of calculations concerns 
Si sublimation experiments. Since Si removal is expected 
to yield an ideally C-terminated surface (C-SiC(001)), which then reconstructs,
we studied the spontaneous reconstruction of an ideally terminated 
C-SiC(001) and the thermal stability of its reconstructed phases.
Starting from the ideal geometry 
vibrating at room temperature, we performed microcanonical MD simulations.
Depending on the initial conditions on ionic positions and velocities,
the ideal geometry transforms either into a c(2$\times$2)  
staggered dimer 
(SD) pattern, or into a 
p(2$\times$1) dimer row (DR) geometry (see Fig.~1).
 At $T=0$, the total energy of the DR is lower than that of the SD
geometry by 0.44 eV/surface dimer\cite{FT5}. 
 We note that the C dimers of the first layer have a
similar bond length in the two reconstructions
(1.36 and 1.37 \AA\ in the DR and SD, respectively),
very close to that of the dimers on the diamond (001) surface\cite{kp95}. 
 In both geometries,
the Si-C distance between the first and second
layers is sensibly longer than its bulk value ($\simeq$5$\%$). 
 However, in the SD reconstruction only the first C layer shows
important deviations from the ideal geometry, whereas 
in the DR reconstruction three
layers take part in the reconstruction, with both the second
and third layer having a large buckling ( 0.2 and 0.37 \AA, respectively). 

 To investigate whether a C-SiC(001) surface can exhibit an SD and/or DR
reconstruction in the temperature regime where 
Si desorption occurs, we studied the
thermal stability of the SD and DR geometries.
 In our calculations, the SD geometry 
is stable against thermal fluctuations up to 1200~K.
 At this temperature the distance between the first and second layer becomes
considerably larger than at $T=0$. 
 Eventually some of the Si-C bonds
break, yielding C dimers bonded only at one end to two Si atoms of the
second layer.
 After some oscillations, these dimers rotate
by about 90 degrees, and are finally bonded to Si atoms at both ends with
one bond per C atom.  Si atoms in the second layer form dimers.
We call such a configuration a bridge (B) pattern (see Fig.~1). 
We also obtained a B reconstruction by deposition of 
C dimers on a Si-SiC(001) surface, as we will discuss below.
Our simulations clearly show that an SD geometry can transform to a B 
geometry at high temperature, whereas no tendency to
form DR configurations was observed within the spanned 
temperature range.

 When heating the DR reconstruction up to 1800~K, we did not observe
any topological change of the surface in spite of
a considerable elongation and occasional breaking of the Si-C bonds 
between the uppermost layers. 
 The C dimers mostly 
oscillate in the surface plane. 
 Dimerization of some Si atoms in the second layer was observed above 1200~K.
 In the DR geometry, 90 degree dimer rotations that could give rise to B
configurations are less favored than in the SD geometry; only a collective
motion of all the dimers could allow such a structural transition.

 In summary, depending on initial conditions, an ideally
terminated C-SiC(001) surface at finite temperature 
can spontaneously reconstruct into two different geometries, which
are both stable up to the temperature range of Si sublimation experiments.
Therefore the reconstruction resulting from Si desorption
is expected to be determined by the kinetics of the sublimation
process and the precise temperature at which Si removal occurs, rather
than by the energetics of the surface at $T=0$.
 Furthermore we find that the coexistence of various reconstruction 
patterns, including DR, SD and B geometries, is possible in the 
temperature range of sublimation experiments\cite{FT10}. 

The photoemission data of
Parril et al.\cite{pc91} for Si sublimated surfaces 
did not reveal the presence of any Si-Si bonds, which would instead be present
in a B geometry. This is consistent with our simulations at $T \leq 1200$ K.
 Our results are also consistent
with the data of Semond et al.\cite{ssmc95}, who obtained rather inhomogeneous 
surfaces by thermal desorption. 
 However there is so far no experimental evidence of p(2$\times$1)
reconstructions resulting from Si sublimation.
 This could be explained either by the kinetics of the
Si desorption process, or by entropic effects at high temperature,
which may favor c(2$\times$2) reconstructions.

 In order to make contact with experiments preparing C-SiC(001) surfaces by 
ethylene deposition on Si-SiC(001) substrates, we
performed a second series of computations. 
 We arranged C dimers in a staggered pattern on an
ideally Si-terminated surface and optimized the geometry of the system. 
The dimers were given two inequivalent (perpendicular) orientations
on the top and on the bottom surfaces.
After annealing, a B reconstruction was obtained on both faces:
the top C dimers rearranged so as to induce
a marked dimerization of the second layer Si atoms.
 The bottom C dimers turned by 90 degrees to yield
eventually the same geometrical pattern as on the top surface. 
 We obtain a remarkably small (1.23 \AA) C-C distance in the C 
dimers, indicating triple bonds,  
and a bond length of the Si dimers (2.37 \AA) similar 
to that in bulk Si (2.35 \AA).
 Contrary to what is found in the SD and DR reconstructions,
the Si-C distance between the uppermost
layers decreases by about $3\%$ with respect to its bulk value.
 As in the DR geometry, three layers take part in the reconstruction. 

 The calculated total energy of the B reconstruction
is lower than that of the SD by 0.18 eV/dimer and higher than that of the DR
by 0.2 eV/dimer. These results agree well with those of Sabisch 
et al.\cite{skmrp96}, although our energy differences are larger than in 
their calculations, and are consistent with the findings of Ref.~\cite{ysj95}.

Our simulations, showing that C$_2$ molecules deposited on a
Si-SiC(001) surface can yield a stable B reconstruction, confirm the 
experimental results of Long et al.\cite{lbr96}, who recently demonstrated 
that C$_2$H$_4$ deposition on Si-SiC(001) lead to a B geometry.
It would be very interesting to explore whether other hydrocarbons 
(containing e.g. two-fold coordinated C atoms)
might lead to the so far unobserved DR reconstruction.
 Powers et al.\cite{ppmhs91} also reported that ethylene deposition
on Si-SiC(001) leads to a B geometry.
 However they suggested that Si desorption
should yield as well a B reconstruction, in spite of difficulties
encountered in fitting
either an SD or a B geometry to their data for Si sublimated surfaces.
 These difficulties can be understood in view of our results,
by a possible coexistence of SD and B
configurations in surfaces prepared by Si desorption.

 We now turn to the discussion of the electronic properties of the C-SiC(001)
surfaces and in particular of surface band structures (SBS).
 In the B reconstruction, the valence band top (VBT), 
located at $\overline{\Gamma}$,
has a clear bulk character, whereas in other parts of
the SBZ the highest occupied state has a surface character.
 This surface state is a
combination of $\sigma$ bonding orbitals on the Si dimers and of $\pi_{\rm z}$ 
bonding orbitals on the C dimers. 
 The conduction band bottom (CBB) (also at $\overline{\Gamma}$)
has again bulk character.
 Above the CBB, we found two surface-bulk resonances. 
 The lowest in energy has large components on the C dimers, given 
by  antibonding combinations of $p_{\rm x}$ orbitals, which are orthogonal 
to the dimer orientation ($\pi^{*}_x$ state). 
 At the SBZ boundaries, a surface state with antibonding 
$p_{\rm z}$ components on C dimers was observed ($p^{*}_z$ state).
 These results are in very good agreement with the spectra of
Long et al.\cite{lbr96}
which show a prominent $p_z$ peak with surface character and 
a less intense $\pi_x$ state at higher energies\cite{FT3}.
 As pointed out in Ref.\cite{lbr96}, the existence of a $\pi_x$ state 
is a clear signature of the B reconstruction.
 Furthermore Long et al. found evidence of states coming from Si-Si bonds
from Si 2p CLS measurements. 
 In our calculation the B structure is semiconducting with a gap of 
$\simeq$~1.4 eV.
 
 The SBS of the SD and DR geometries are simpler than 
that of the B reconstruction, due to the absence of Si-Si bonds.
In both surfaces the highest occupied (HO) and 
lowest unoccupied (LU) states at the $\overline{\Gamma}$ point
have bulk character.
Close to the VBT and CBB, three surface states were observed: a $\pi_z$, a 
$\pi^{*}_y$ and a $\pi^{*}_z$ state, localized on the C dimers\cite{FT7}. 
Our SBS results agree very well with those reported 
by Sabisch et al.\cite{skmrp96}.

 In order to predict the results of STM experiments for the C-SiC(001)
surfaces, we report in Fig.~2
the square moduli of the HO and LU surface states for the three geometries,
on a plane parallel to the surface at a distance of 1.5 \AA.
There is an unequivocal difference between the images of the 
p(2$\times$1) and the c(2$\times$2) surfaces.
 Drawing a clear distinction between the B and SD images is more delicate.
 When rotated by 90 degrees,
the images of the $\pi^{*}_x$ and $\pi_z$ states of the B surface
may not be clearly distinguishable from those of the
$\pi^{*}_y$ and $\pi^{*}_z$ of the SD surface.
 The key difference between the two sets of images lies in the state occupation:
the $\pi$ states perpendicular to the surface are occupied (empty) 
and those parallel to the surface are empty (occupied) in the B (SD) geometry.
 Fig.~2 suggests that STM experiments
could be effective probes to identify the various reconstructions,
and possibly reveal the coexistence of 
domains with 
different reconstructions on the same surface, as was recently
observed in diamond (001)\cite{KSS95}.

 We further analyzed the differences between the 
reconstructions of C-SiC(001), by computing Si-2p and C-1s core level shifts 
with respect to bulk SiC. We adopted the procedure of Ref.~\cite{phc96}
and ionized atoms of the two uppermost layers\cite{FT4}.
The values obtained for Si-2p CLS in the
SD and DR reconstructions are similar (-0.12~eV and
-0.09~eV), but different in magnitude and sign from that calculated 
for the B reconstruction (0.32~eV). Indeed, Si-Si bonds only exist 
in the B geometry.
C-1s CLS of the atoms of the first layer
are instead very similar in the three structures
(-0.44 and -0.48 and -0.46 eV for the SD, DR and B geometries,
respectively), although only in the B reconstruction C dimers are triply bonded.
 Our calculated C-1s CLS for the B geometry
is smaller than the experimental value\cite{lbr96} (-1.1 eV),
the error being close to that found in a variety
of carbon systems\cite{prs93}. Whereas the LDA is believed to give
a good quantitative description of CLS involving Si atoms\cite{phc96},
only a qualitative agreement between theory and experiment
is obtained for CLS of several C systems\cite{prs93}. 
For the B geometry we also computed the CLS of C atoms of inner layers.
These CLS have magnitudes similar to those of the uppermost C atoms
but have opposite signs. We therefore expect the presence
of broad peaks in C-1s core absorption data.

{\it Si-terminated surfaces} --
In our calculations for Si-SiC(001), the ground state geometry 
at $T=0$ is a p(2$\times$1) dimer row reconstruction,
which lies a few meV/atom lower than the ideal surface geometry.
The Si dimer bond length ($d_{\rm Si}$) is 2.58~\AA, 
larger than the
value (2.31~\AA) fitted to LEED data\cite{pmhs92}.
 We investigated the thermal stability of the reconstructed surface
by performing constant temperature MD simulations in the range 
200 K $\leq T \leq$ 800~K.
At $T \simeq$ 300~K the surface dimer length varies between 
$\simeq$ 2.3 and 2.9 \AA, i.e.\ close to the value of the 
ideal structure (3.03 \AA).
 At higher temperatures (300~K $\leq T \leq$ 500~K), the 
surface oscillates between a p(2$\times$1) dimer row reconstruction
and an ideal surface geometry.
 The large thermal fluctuations of $d_{\rm Si}$
could be responsible for the difficulties 
encountered in achieving a good fit of LEED 
data for Si-SiC(001) surfaces\cite{k89,pmhs92,s96}.
 At $T \geq$ 500K, we observed a tendency to form a c(4$\times$2)
reconstruction, where neighboring dimers have opposite buckling\cite{further}.
 The ability of the Si-SiC(001) surface to sample several
reconstructions in a relatively small temperature range
could explain the differences between
recent\cite{s96} and older LEED experiments\cite{k89,pmhs92}.
 Finally, we evaluated Si-2p CLS and found
a value (0.85~eV) which is in very good agreement with
experiment (0.8-1.0 eV)\cite{s96}.

 In summary we investigated the effect of preparation conditions on
C-terminated $\beta$-SiC (001) surfaces, as well as the thermal stability
of both C- and Si-terminated surfaces.
 Computations of band structures and  
C-1s and Si-2p core level shifts of the stable C-SiC(001)
geometries point at STM images and Si-2p CLS
as effective probes to identify the different reconstructions.

 We thank A.~Pasquarello and M.~Sabisch for useful discussions. 
The calculations were performed on the NEC SX3/4 of CSCS in Manno, Switzerland.
Work supported by the Swiss NSF Grant No.~20-39528.93 (GG) and the italian
CNR (AC) through a FNS-CNR cooperation. 

\begin{figure}
\caption{
Top view of the dimer row (DR),
bridge (B) and staggered dimer (SD) reconstructed geometries
of the C-SiC(001) surface. Three layers are shown.
Black and white spheres indicate C and Si atoms. Axes are oriented as
in Ref.~[11].  In the three geometries dimers are not buckled.
}
\end{figure}

\begin{figure}
\caption{
Square moduli of the highest occupied (lower panel) and lowest unoccupied
(upper panel) surface states in a plane parallel to the surface,
at a distance of 1.5 \AA,
for the DR (left, $\pi^{*}_y$ and $\pi^{*}_z$ states),
the B (center, $\pi_z$ and $\pi^{*}_x$ states) and the SD
(right, $\pi^{*}_y$ and $\pi^{*}_z$ states) reconstructions of the
C-SiC(001) surface (see Fig.~1).
White bars indicate C dimers.
In the B reconstruction, at 1.5 \AA\  from the surface
the electronic states are dominated by C dimers components 
and Si dimers components are not visible.
}
\end{figure}

\begin{thebibliography}{10}
\bibitem{convergence}
Geometry optimizations were considered converged when
the largest ionic force was smaller than 0.005 eV/\AA.
After optimization, residual stresses on the unit cells
varied from 0.002 to 0.032 Mbar,
to be compared with the computed bulk modulus of cubic SiC
(2.17 Mbar). The experimental bulk modulus is 2.24 Mbar.
\bibitem{FT5}
Total energy differences obtained with the 11 and 19 layer slabs
differ by $\simeq$ 0.03 eV/dimer. Eigenvalue differences involving
surface states are very similar in the 11 and 19 slab calculations,
whereas those involving bulk states can differ by as much as 0.5 eV.
\bibitem{FT10}
 Graphitization could also take place after desorption at high T, given the
 number of Si-C broken bonds in our simulation for the SD geometry at 1200~K.
 Since the mismatch between the graphite and SiC lattice constants is large,
 graphitization is likely to proceed by formation of microclusters and not of
 a long range two dimensional lattice. However such a process cannot be
 observed in cells containing only 8 atoms per layer.
\bibitem{FT7}
 In our calculation both the DR and SD surfaces 
 are metallic, as a result of a
 small overlap between the $\pi^{*}_y$ and $\pi^{*}_z$ surface bands. For
 example, in the SD reconstruction the $\pi^{*}_y$ ($\pi^{*}_z$) state is
 occupied (empty) in most of the SBZ. This band overlap could be an artifact
 of the LDA, as pointed out in Ref.~15. In order to estimate the effect of 
 band occupation on MD simulations,
 we forced full occupation of the $\pi^{*}_y$ 
 all over the SBZ in the SD reconstruction. 
 The total energy thus obtained is higher than that of
 the ground state by 0.05 eV/dimer and the dimer bond length varies from
 1.37 to 1.39 \AA.
\bibitem{FT4}
 The convergence with $E_{\rm cut}$ of C-1s CLS was tested by
 computing relative core shifts in C$_2$H$_4$ and C$_2$H$_2$. We obtained 0.60
 and 0.57 eV with $E_{\rm cut}$=40 and 50 Ry, respectively.
 Size effects on C-1s and Si-2p CLS
 were tested by comparing values obtained with the 11 and the
 19 layer slabs, which differed by $\simeq$ 0.05 eV.
\bibitem{further}
A more detailed analysis will be given in a forthcoming publication.
\end{thebibliography}
\end{document}